\documentclass[aps,pre,twocolumn,groupedaddress,showpacs,floatfix]{revtex4}
\usepackage{graphicx}
\usepackage{pst-all}
\usepackage{color}
\usepackage{amsmath}
\usepackage{textcomp}
\newcommand{\comment}[1]{}

\def\Ctwo{$\mathcal{C}_2$\ {}}
\def\Cone{$\mathcal{C}_1$\ {}}
\def\Na{N_f^{(a)}}
\def\Nn{N_f^{(n)}}

\begin{document}
\title{The properties of attractors of canalyzing random Boolean networks}
\author{U. Paul, V. Kaufman, B. Drossel}
\affiliation{Institut f\"ur Festk\"orperphysik,  TU Darmstadt,
Hochschulstra\ss e 6, 64289 Darmstadt, Germany }
\date{\today}
\begin{abstract}
We study critical random Boolean
networks with two inputs per node that contain only canalyzing functions.
We present a phenomenological theory that explains how a frozen core of nodes that are frozen on all
attractors arises.
This theory leads to an intuitive understanding of the system's dynamics as it demonstrates the
analogy between standard random Boolean networks and networks with canalyzing functions only.
It reproduces correctly the scaling of the number of nonfrozen nodes with system size.
We then investigate numerically attractor lengths and numbers, and explain the findings in
terms of the properties of relevant components. In particular we show
that canalyzing networks can contain very long attractors, albeit
they occur less often than in standard networks.
\end{abstract}
\pacs{89.75.Hc, 05.65.+b,  02.50.Cw}
\maketitle

\section{Introduction}
\label{sec:intro}

Random Boolean networks are often used as generic models for the
dynamics of complex systems of interacting entities, such as social
and economic networks, neural networks, and gene or protein
interaction networks \cite{kauffman:random}. The simplest and most
widely studied of these models was introduced in 1969 by Kauffman
\cite{kauffman:metabolic} as a model for gene regulation.  The system
consists of $N$ nodes, each of which receives input from $K$ randomly
chosen other nodes. The network is updated synchronously, the state of
a node at time step $t$ being a Boolean function of the states of the
$K$ input nodes at the previous time step, $t-1$.  The Boolean
update functions are randomly assigned to every node in the network,
and together with the connectivity pattern they define the \emph{realization}
of the network. For any initial condition, the network eventually
settles on a periodic attractor.  Of special interest are
\emph{critical} networks, which lie at the boundary between a frozen
phase and a chaotic phase \cite{derrida:random,derrida:phase,moreira:canalcritcond}.  In the
frozen phase, a perturbation at one node propagates during one time
step on an average to less than one node, and the attractor lengths
remain finite in the limit $N\to \infty$. In the chaotic phase, the
difference between two almost identical states increases exponentially
fast, because a perturbation propagates on an average to more than one
node during one time step \cite{aldana-gonzalez:boolean}.

Critical networks with $K=2$ inputs per node have been studied by a
variety of authors. A $K=2$ network is critical if \emph{frozen} and
\emph{reversible} update functions are chosen with equal
probability. The remaining update functions are \emph{canalyzing},
i.e., one input can fix the output of a node, irrespective of the
value of the second input. Table \ref{tab1} shows the 16 update
functions of $K=2$ networks. 
\begin{table} \begin{center}
\begin{tabular}{|c||c|c||c|c|c|c||c|c|c|c|c|c|c|c||c|c|}\hline In&
\multicolumn{2}{|c||}{$\mathcal{F}$}&
\multicolumn{4}{|c||}{${\mathcal{C}}_1$}&
\multicolumn{8}{|c||}{${\mathcal{C}}_2$}&
\multicolumn{2}{|c|}{$\mathcal{R}$}\\\hline
00&1&0&0&1&0&1&1&0&0&0&0&1&1&1&1&0\\
01&1&0&0&1&1&0&0&1&0&0&1&0&1&1&0&1\\
10&1&0&1&0&0&1&0&0&1&0&1&1&0&1&0&1\\
11&1&0&1&0&1&0&0&0&0&1&1&1&1&0&1&0\\\hline \end{tabular} \end{center}
\caption{The 16 update functions for nodes with two inputs. The first
column lists the 4 possible states of the two inputs, the other
columns represent one update function each, falling into the
classes \emph{frozen} ($\mathcal{F}$), \emph{canalyzing} (${\mathcal{C}}_1$ and ${\mathcal{C}}_2$) and \emph{reversible} ($\mathcal{R}$).}  \label{tab1} \end{table}
Critical networks that contain a nonvanishing proportion of frozen and
reversible update functions are in the meantime relatively well
understood, see
\cite{bilke:stability,socolar:scaling,samuelsson:superpolynomial,gershenson:updating,klemm:asynchrstability,drossel:onnumber,kaufmanandco:scaling}%
.  They contain three
classes of nodes, which behave differently on attractors.  First,
there are nodes that are frozen on the same value on every
attractor. Such nodes give a constant input to other nodes and are
otherwise irrelevant. They form the \emph{frozen core} of the
network. Second, there are nodes without outputs and nodes whose outputs go only to irrelevant
nodes. Though they may fluctuate, they are also classified as
irrelevant since they act only as slaves to the nodes determining the
attractor period. Third, the \emph{ relevant nodes} are the nodes
whose state is not constant and that control at least one relevant
node. These nodes determine completely the number and period of
attractors. If only these nodes and the links between them are
considered, these nodes form loops with possibly additional links and
chains of relevant nodes within and between loops. We call a set of
relevant nodes that are connected in this way a \emph{relevant
component}. The nonfrozen nodes that are not relevant sit on trees
rooted in the relevant components. In \cite{socolar:scaling}, it was
found that the number of nonfrozen nodes scales in the limit
$N\to\infty$ as $N^{2/3}$ and the number of relevant nodes as
$N^{1/3}$. This result was confirmed by an analytical calculation in
\cite{kaufmanandco:scaling}, where it was additionally shown that
the number of nonfrozen nodes with two nonfrozen inputs scales as
$N^{1/3}$, and that the number of relevant nodes with two relevant
inputs remains finite in the limit $N\to\infty$. The mean number of
relevant components was found to be proportional to $\ln N$, and all
but the largest relevant components are simple loops.

Canalyzing networks share many features of other critical
networks. Thus, the calculation by Samuelsson and Troein
\cite{samuelsson:superpolynomial} of the number of attractors can be
generalized to canalyzing networks \cite{drossel:onnumber}, implying
that canalyzing networks also have of the order of $N^{2/3}$ nonfrozen
nodes and at most $N^{1/3}$ nonfrozen nodes with two relevant inputs,
and that the mean attractor number increases faster than any power law
with the network size.  Whether the nonfrozen nodes are the same on
all attractors in canalyzing networks (as is the case of networks
containing frozen update functions), can not be decided from previous
work. In particular, the detailed results of
\cite{kaufmanandco:scaling} can only be derived if there are nodes
with frozen functions. For these reasons, a separate study of
canalyzing networks is needed. It is the main aim of this paper to
show how the attractors and the frozen nodes arise in canalyzing
networks. We will see that canalyzing networks also have a frozen
core, which means that most frozen nodes are the same on all
attractors. It follows then that all results of
\cite{kaufmanandco:scaling} about the relevant part of the network can
be applied also to canalyzing networks.  We will also put an end to
the long-standing belief that canalyzing networks have less and
shorter attractors. These features were argued to make canalyzing
networks biologically more relevant \cite{kauffman:random}.

Let us therefore focus on $K=2$ networks that contain only
\Ctwo functions. These functions take one value (0 or 1) three times
and the other one once. This means that each of the two inputs can fix
the output of the function irrespective of the other input. For
instance, the output of the first \Ctwo function shown in Table
\ref{tab1} must be 0 if the first input is 1, irrespective of the
second input. It must also be 0 if the second input is 1, irrespective
of the first input.  Each of the 8 \Ctwo functions is chosen with
equal probability in our simulations. We will compare our results for
\Ctwo networks with those of standard random Boolean networks (RBNs),
where all 16 update functions have the same weight. Part of
our results will also be compared to those of \Cone networks, where
the update functions are chosen only from the \Cone class. The \Cone
networks can be trivially mapped on critical $K=1$ networks by
removing the link to the input to which the node does not
respond. These networks have no frozen core. They have of the order of
$\sqrt N$ relevant nodes, arranged in $\sim \ln N$ simple loops (see
\cite{drossel:number}), with the largest loop length being of the
order $\sqrt{N}$. The other nodes sit on trees rooted in these loops.

In the next Section, we study numerically the frozen nodes in order to
find out if the same nodes are frozen on all attractors of \Ctwo
networks. In Section \ref{theory}, we explain the results of the numerical simulations using
phenomenological arguments and analytical calculations.  In Section
\ref{attractors}, we study the number and length of attractors of
\Ctwo networks and compare the results to those of other network types.
\comment{Again, we use phenomenological arguments to explain the results.
^We can do without this sentence^, since otherwise the latex-output is ugly, we need some space...}
Finally we summarize and discuss our results in the last Section.

\section{The frozen core}
\label{sec:froz}

From a generalization of the work of Samuelsson and Troein
\cite{samuelsson:superpolynomial} to all critical $K=2$ random Boolean
networks \cite{drossel:onnumber}, we know that for canalyzing networks the
number of nonfrozen nodes scales for large network size in the same
way as for RBNs, i.e., with $N^{2/3}$. In RBNs, the nodes frozen on
all attractors (i.e., the nodes belonging to the frozen core) can be
identified by starting with the nodes with frozen update functions and
by iteratively determining nodes that become frozen because of frozen
inputs, see \cite{kaufmanandco:scaling}. In canalyzing networks
there are no frozen functions to start with, and this method cannot be
applied. It therefore arises the question whether canalyzing networks
have a frozen core at all, or whether different attractors have
different nonfrozen nodes.  In the following, we will answer this
question using computer simulations.  
\begin{figure}
\begin{center}
\vskip 0.5cm
\includegraphics*[width=0.95\columnwidth]{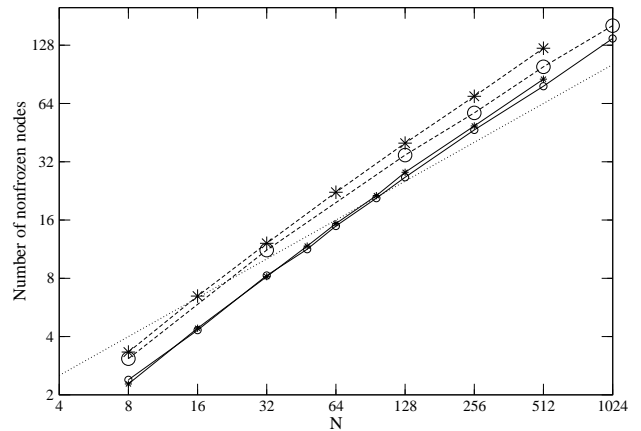}
\end{center}
\caption{Mean number of nonfrozen nodes for canalyzing \Ctwo networks
(stars) and RBNs (circles). Solid lines connect the data
points for $N-N_f^{(a)}$ (mean number of nonfrozen nodes per
attractor), dashed lines the data points for $N-N_f^{(n)}$ (mean number
of nodes that are nonfrozen at least on one found attractor). The
dotted line is a power law $N^{2/3}$.  Different attractors are
counted only once, without considering their basins of attraction. We
have considered 1000 initial states per network, averaged over 2000
networks.}
\label{fig:froz}
\end{figure}

In Fig. \ref{fig:froz} we show the average number $\Na$ of frozen nodes
per attractor, both for canalyzing networks and for RBNs. We actually plot the difference $(N-\Na)$ as a function of $N$ in order to better see the asymptotic behavior. The other two curves show the
difference $(N-\Nn)$, where $\Nn$ is the
number of nodes frozen on all attractors found in the simulation of a network. The technical details can be found in the caption to the figure.

One can clearly see the similarity of the results for the mean number
of nonfrozen nodes per attractor $(N-\Na)$ for canalyzing networks and
for RBN, in agreement with
\cite{samuelsson:superpolynomial,drossel:onnumber}. The expected power
law $N^{2/3}$ is not yet reached for the system sizes used in the
simulation and is only approached slowly with increasing $N$. The
number of nonfrozen nodes in the simulations was only of the order of
100 for the largest simulated networks, which is yet too small to see
the asymptotic behavior. (And the number of relevant nodes, which increases
as $N^{1/3}$, is only of the order $10$ for the largest simulated
networks.)

Our results for  $(N-\Nn)$ suggest that canalyzing networks have a frozen core and of the order of $N^{2/3}$ nonfrozen nodes, because the curves for $(N-\Nn)$ differ only by a constant factor from those for  $(N-\Na)$ for both network types. Furthermore, there are of the order of $N^{2/3}$ nodes that are frozen only on part of the attractors. The factor between the two curves is larger for canalyzing networks than for RBN. 

In the following, we will explain the reason for the constant factor between the curves for  $(N-\Na)$ and  $(N-\Nn)$. Since this point has not yet been discussed for RBNs, we will consider both RBNs and canalyzing networks. The explanation of the origin of the frozen core in canalyzing networks will be postponed until the next Section.

The difference between the curves for $(N-\Na)$ and $(N-\Nn)$ is due
to those nodes that are frozen on some attractors, but not on all
attractors. These nodes do not belong to the frozen core, and they are
therefore relevant nodes or sit on nonfrozen trees that are rooted in
relevant components. In Section \ref{sec:intro}, we have mentioned that most
relevant components consist of simple loops, and that only a
few large components are more complex and contain relevant nodes that
have two relevant inputs. Clearly, since the dynamics of the nonfrozen
nodes in the trees is determined by the dynamics of the relevant
nodes, all nodes of a relevant component and the nonfrozen trees
rooted in it undergo a cycle of the same period (when the attractor
has been reached), which is determined by the initial state of the
relevant nodes of that component. If this cycle has period one, all
nodes of this component are frozen on this attractor. We therefore
have to show that a finite fraction of all nonfrozen nodes are on
cycles of length 1 on a finite fraction of all attractors. This would lead
to a constant factor between the curves for $(N-\Na)$ and $(N-\Nn)$.

Let us first consider relevant components that are simple loops, and
their nonfrozen trees. The mean number of relevant loops of length $l$
is $1/l$ for all $l$ up to a cutoff $l_c \sim N^{1/3}$.  The mean
size of a tree rooted in a relevant node is $N^{1/3}$. The largest of
these components consists therefore of the order of $N^{2/3} $ nodes
(including the nonfrozen trees), and if such a component reaches a
fixed point attractor with nonzero probability, we have explained the
factor between the two curves. A relevant loop of length $l$ has
either two fixed points (if the loop is ``even'', i.e. if the state of
a node is repeated after $l$ time steps) or none (if the loop is
``odd'', i.e. the state of a node is inverted after $l$ time
steps). Each case occurs with probability 1/2. The number of
attractors of a component with a loop of length $l$, however,
increases exponentially with $l$, and for this reason only a vanishing
proportion of attractors of components of a size of the order $l_c$
are fixed points in the limit of large $N$. 

Next, we consider complex relevant components. In contrast to simple
loops, where each initial state is part of a periodic cycle in state
space, more complex components can have fixed points that are true
attractors, i.e. that are reached from a nonvanishing proportion of
initial states (but not from all initial states). One example of such
a component in RBNs was discussed in \cite{kaufman:on}. It is a loop
with an additional chain of nodes within the loop, such that there is
one node that has two relevant inputs. From
\cite{kaufmanandco:scaling}, we know that this component occurs with
nonvanishing probability in an RBN. In the case that the update
function of the node with two inputs is 0 only if both inputs are 0
and that the two numbers of nodes on the two subloops have a common
divisor greater $1$, all apart from a finite number of initial
conditions end up on the same fixed point. The existence of such
components does not only explain the multiplicative factor between the
curves for $(N-\Nn)$ and $(N-\Na)$, but it explains also why this
factor is larger for \Ctwo networks than for RBNs. The probability
that an update function as in the above example is chosen at the
relevant node with two inputs is larger for canalyzing networks.

\begin{figure}
\begin{center}
\vskip 0.5cm
\includegraphics*[width=0.95\columnwidth]{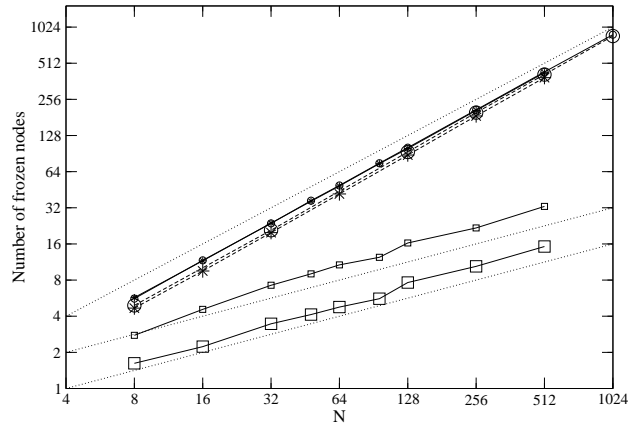}
\end{center}
\caption{Number of frozen nodes for $K=1$ networks. Large squares: $N_f^{(n)}$; small squares: $N_f^{(a)}$. For comparison, the corresponding curves for RBNs and \Ctwo networks are also shown, the data is taken from Fig. \ref{fig:froz}. The dotted lines correspond to the power laws $N,\; \sqrt{N}$ and $\sqrt{N}/2$.}
\label{fig:froz2}
\end{figure}

We conclude this Section by comparing \Ctwo networks with \Cone networks,
which are canalyzing networks, but with update functions of the \Cone
class.  Our simulation results for $\Na$ and $\Nn$ are shown in
Fig. \ref{fig:froz2}.  Both curves increase for \Cone networks as
$\sqrt N$. Asymptotically, $\Nn$ is of the order $\sqrt{N}/2$, since
only even loops of length $1$ can be frozen on all attractors, while
the average tree size is of the order of $\sqrt{N}$. To calculate
$\Na$ we note that only even loops of length $l$ have (two) fixed
points, and that they do not reach these fixed points for $2^l-2$
 initial conditions. The contribution to the average number of
frozen nodes per attractor from simple loops of length $l$ is then
$1/l\cdot 1/2\cdot 2l/2^l\cdot \sqrt{N}$, where we again take the
probability of an even loop $1/2$ and the average size of a tree of
the order of $\sqrt{N}$ into account. Summing over $l$ leads to
$\Na\to \sqrt{N}$. We conclude that, differently from \Ctwo networks,
the ratio $(N-\Nn)/(N-\Na)$ approaches 1 asymptotically. As we have learned,
the reason for a larger factor between the two curves for RBNs and
\Ctwo networks is the existence of complex relevant components. In
\Cone networks all relevant components are simple loops.

\section{Self-freezing loops}
\label{theory}

In this Section, we want to explain how the frozen core arises in
\Ctwo networks and find some of its properties. We also estimate its
size by means of analytical arguments. The results are in agreement
with \cite{samuelsson:superpolynomial,drossel:onnumber}, confirming our intuitive
understanding of the origin of the frozen core. 

Since a \Ctwo network has no nodes with a frozen function, the frozen
core cannot be formed starting from single frozen nodes. Instead,
there must exist groups of nodes that fix each other's value and do
not respond to changes in nodes outside this group.  Let us consider
the simplest example of such a group, namely a loop where each node
canalyzes (fixes) the state of its successor once it settles on its
majority bit (the one occuring 3 times in its update function table).
We call such loops \emph{self-freezing loops}.  In the following, we
first discuss these self-freezing loops, before discussing how a
frozen core that contains almost all nodes can be formed starting from
these loops.

If all nodes of a self-freezing loop are on their majority bits, it
stays frozen forever.  Starting from an arbitrary initial state, the number of
nodes of  a self-freezing loop on majority bits cannot decrease with time, since each such node
drives its successor to its majority bit.  It remains constant only in
the unlikely case that all inputs from outside the loop to the nodes
of the loop are fixed on the noncanalyzing value. We can therefore
assume that self-freezing loops are usually frozen on all attractors,
at least if the loops are large.  As we will see, most nodes that are
part of self-freezing loops sit in loops with a size of the order of
$N^{1/3}$.

The number of nodes on self-freezing loops can be estimated as
follows. The probability that a given node constitutes a self-freezing
loop of length $1$ is $1/N$ for a network with $N$ nodes. It is the
product of the probability $2/N$ that the node is self-connected and
the probability $1/2$ that the node is canalyzed by its own majority
bit. There is thus on average one self-freezing loop of length $1$ per
network. With the same line of reasoning, the average number of
self-freezing loops of length $2$ per network is obtained to be
$\binom{N}{2}\left(\frac{2}{N}\right)^2\left(\frac{1}{2}\right)^2\approx
\frac{1}{2}$. For the self-freezing loops of length $l>2$ one has to
take into account an additional factor, corresponding to the number of
ways to construct a directed loop out of $l$ nodes. The number of
self-freezing loops of length $l$ per network is found to be
$1/l$. The overall number of nodes on self-freezing loops $f_0$ is
then
\begin{equation}
\label{eq:f0}
f_0 = \Sigma_{l=1}^{l_c}\frac{1}{l} l = l_c,
\end{equation}
$l_c$ being the cutoff in loop length. This simple probabilistic
considerations applies if $l_c$ is much smaller than $N$.

We can obtain a confirmation of this estimation and a result for the
value of $l_c$ by mapping the problem of finding a self-freezing loop
in a \Ctwo network onto the problem of finding the relevant nodes
sitting on relevant loops in a critical network that contains no canalyzing
functions at all, but only $\mathcal{R}$ and $\mathcal{F}$
functions. Whether a randomly chosen node in such a network is part of a
relevant loop is determined by the following algorithm. Consider the
two inputs of this node. With probability 1/4, both inputs have a
frozen update function, and the node is not relevant. With probability
1/2, one input has a frozen update function and the other one a reversible
one. In this case we draw a link to this reversible input node and thus mark it for
investigation of its two inputs in the next step.
With probability 1/4, both inputs have
reversible update functions, and we draw links to both of them. We
iterate this procedure, choosing at each step the two inputs of a node
at random from all nodes, and drawing links to those inputs that do not have frozen update function. The procedure continues until we either
find a connection back to the original node (in which case it is
relevant), or until no more links can be drawn (in which case the
original node is not part of a relevant loop). From the results of our
article \cite{kaufmanandco:scaling}, we know that there is a mean number of
$1/l$ relevant loops of size $l$ in such a network, and that the
cutoff in the size of relevant components scales as $N^{1/3}$.

Now, we turn to the procedure of finding self-freezing loops in
\Ctwo networks and show that it is identical to the procedure just
described. We start with a randomly chosen node and determine whether
it is part of a self-freezing loop. Consider the two inputs of this
node. With probability 1/4, the majority bit of neither input
canalyzes the chosen node, and the node is not on a self-freezing
loop. With probability 1/2, the majority bit of one input does not
canalyze the chosen node, and the majority bit of the other input
canalyzes it. Let us draw a link to this input node and consider its
two inputs in the next step. With probability 1/4, the majority bits
of both inputs canalyze the chosen node, and we draw links to both of
them. We iterate this procedure, choosing at each step the two inputs
of a node at random from all nodes, and drawing links to those inputs,
whose majority bits canalyze the node. The procedure continues until
we either find a connection back to the original node (in which case
it is part of a self-freezing loop), or until no more links can be
drawn (in which case the original node is not part of a self-freezing
loop). The analogy of the two procedures is obvious, and we conclude
$l_c \sim N^{1/3}$ and $f_0 \sim N^{1/3} $.

Obviously, nodes depending on or canalyzed by the frozen nodes of the
self-freezing loops freeze also, and such nodes may lead to the
freezing of further nodes, etc. We introduce a dynamical process in
order to determine the total number of nodes that become frozen
because of the self-freezing loops.  We denote with $f$ the number of
nodes that have already become frozen during the process, and the
influence of which on other nodes has yet to be determined.  $n_1$ is
the number of nodes for which we already know that one of their inputs
is frozen but does not canalyze them, and $n_2$ is the number of nodes
for which no frozen input was yet identified.  Initially, $n_1=0$, $f
= f_0$, $n_2=N-f_0$ and $n\equiv f+n_1+n_2=N$.  We answer for one of
the frozen nodes at a time the question whose input it is. It is an
input to any of the $n_2$ nodes with probability $2/n$.  With equal
probability $1/2$, the node either becomes frozen by this input, or it
becomes a non-frozen node with effectively one input (called a \Cone
node in the following). If a node with one input chooses the given
frozen node as input (that happens with probability $1/n$), it becomes
frozen. At each step, the connected frozen node is being excluded from
further consideration. The dynamical process stops when all the nodes
are frozen (which is improbable) or when there are no more frozen
nodes the influence of which on other nodes has not yet been
determined.  The dynamical equations for this process are
\begin{align}
\label{eq:core_dynam}
\nonumber
\Delta f &= -1+\xi_1+\xi_2\,,\\
\Delta n_1 &= (n_2-n_1)/n\,,\\\nonumber
\Delta n_2 &= -2 n_2 /n\,.
\end{align}
The stochastic Poisson distributed noise terms $\xi_1$ and $\xi_2$
with the mean values $n_1/n$ and $n_2/n$ respectively have to be taken
into account in the equation for $f$, since $f$ becomes small during
the process, so that the noise becomes important, compare
\cite{kaufmanandco:scaling}. The sum $n=f+n_1+n_2$
decreases by 1 in each step.

Simulations of this process show that the total number of nodes that
are frozen because of the self-freezing loops is around $\sim
N^{0.8}$, and that the number or nodes that are not fixed by the
self-freezing loops is of the order of $N$. The number of nodes frozen
because of the self-freezing loops is not large enough to explain the
simulation data of the previous Section. We therefore have to find a
mechanism that generates more frozen nodes. This is found by extending
the definition of \emph{self-freezing loops}.  We have just seen that
nodes with one nonfrozen input appear as we identify frozen
nodes. Among the nodes that are not frozen by the original
self-freezing loops, there are new types of self-freezing loops that
contain chains of nodes with one nonfrozen input between \Ctwo nodes.
If a chain between two \Ctwo nodes as a whole inverts its input, the
inverted majority bit of the first \Ctwo node has to canalyze the
second \Ctwo node. As with original self-freezing loops we can claim
that the generalized self-freezing loops are usually frozen on
attractors.  At the end of the process described by
(\ref{eq:core_dynam}), the generalized self-freezing loops need to be
found. The only effect of nodes with \Cone functions in such loops is
to delay the signal propagation between two adjacent \Ctwo nodes. The
remaining $n_2$ nodes with \Ctwo functions can therefore be considered
as an effective \Ctwo network, which leads to ${n_2}^{1/3}$ nodes
on generalized self-freezing loops with similar loop size statistics
as discussed above. The independence of the second
search for self-freezing loops from the first one is due to the large
enough number $n_1$ at the end of the process
(\ref{eq:core_dynam}). This $n_1$ ensures that typical self-freezing
loop in the effective \Ctwo network have insertions of \Cone chains.

With the new self-freezing loops we again run the dynamical process of
type (\ref{eq:core_dynam}).  We can even take over the values of
$n_1$, $n_2$ and $n=n_1+n_2$ at the end of the first process, since
${n_2}^{1/3}$ frozen nodes are negligible in comparison with
$n_2$. Therefore the two equations
\begin{align}
\label{eq:core_dynam_overall}
\nonumber
\Delta n_1 &= (n_2-n_1)/n\,,\\
\Delta n_2 &= -2 n_2 /n\,.
\end{align}
apply to both processes together. The equation for $n$ is
$\Delta n = -1$, as before. The solution of these equations is obtained by going to differential equations for $dn_1/dn$ and $dn_2/dn$, which have the solution
\begin{align}
\label{eq:solution_ni}
n_2 = \frac{n^2}{N}\,,\\
n_1 = n-\frac{n^2}{N}\, .
\end{align}
In the same way, at the end of the second process we have again an
effective \Ctwo network, with chains containing newly generated \Cone
nodes.  The number of remaining \Cone nodes increases in the second
process, the number of \Ctwo nodes decreases, thus leading to an
increasing weight of \Cone nodes in the nonfrozen network.  Equations
(\ref{eq:core_dynam_overall}) are now applied to a third process,
similar to the first two.

\comment{
\emph{This is an obsolete  comment, please let it be here for my memoirs.} We could run the process differently:
find one single sf-loop, determine the frozen nodes due to it, find the next one, determine nodes,
getting frozen then, and so on. Then, seemingly, the last two equations in (\ref{eq:core_dynam}) apply.One could argue, this way of introducing the process requires less explanations and is therefore
``better''. But here you still have to keep track of the number of \Cone nodes produced,
since the applicability of the last two eqs. in (\ref{eq:core_dynam}) relies on the fact
that the number of nodes used
to initiate each consequent process of type (\ref{eq:core_dynam}) is negligible in comparison
to current values $n_1$ and $n_2$ themselves. Additionally, introducing the outer process in the
mentioned way, we can't learn much about the topology of the frozen core (loops lengths distribution).
\emph{End of comment.}
}

The repeated process of identifying generalized self-freezing loops
and the nodes frozen by them breaks down when the remaining nonfrozen
nodes cannot be considered as an effective \Ctwo network any
more. This happens when the proportion $n_2/n$ of \Ctwo nodes becomes
of the order $\sim 1/\sqrt{n}$. Then, in the process of building a
self-freezing loop, there occur \Cone chains of an average length of
the order $\sim \sqrt{n}$ between \Ctwo nodes. Now, the probability to
attach a \Ctwo input at the end of the chain is of the same order of
magnitude $1/\sqrt{n}$ as the probability to attach some node of this
chain at the end of the chain, in which case the chain becomes a loop,
and the assembly of self-freezing loop becomes improbable.
\comment{
A more elaborate estimation of the proportion $n_2/n$ where \Cone loops arise as a typical input of each \Ctwo node yields the same result.
}

Let us denote by $N_{nf}$ the average number of nonfrozen nodes in
\Ctwo networks and by $N_2$ the average number of nonfrozen nodes with
two inputs. $N_1=N_{nf}-N_2$ is then the average number of nodes with
one nonfrozen input. The breakdown condition for the iterated process
becomes then $N_2\sim \sqrt{N_{nf}}$. Inserting the condition
$n_2 \sim \sqrt{n}$ in the solution (\ref{eq:solution_ni}), we obtain
\begin{align}
\label{eq:nonfroz}
N_{nf} & \sim N^{2/3}\,,\\\nonumber
N_2 & \sim N^{1/3}\,.
\end{align}
This is in agreement with the results of
\cite{samuelsson:superpolynomial,drossel:onnumber} and confirms our intuitive
understanding of the frozen core. The frozen core consists of
self-freezing loops, which arise in the interated process described in
this Section. The number of nonfrozen nodes is of the order $N^{2/3}$,
with only $N^{1/3}$ nonfrozen nodes having 2 nonfrozen inputs. The
properties of the nonfrozen part of the network are therefore the same
as those of RBNs, and we can take over the results obtained for the
nonfrozen part of RBNs. In particular, we can conclude that the number
of relevant nodes scales as $N^{1/3}$, with only a finite number of
them having two nonfrozen inputs, and with most relevant components
being simple loops. 

The considerations of this Section can be repeated without change also for
mixed \Cone and \Ctwo critical Boolean networks, consisting of nonfrozen nodes
 with one input and of nodes with two inputs having update functions of class \Ctwo,
provided that the number of nodes with two inputs is larger than $\sqrt{N}$ (otherwise we are left with a \Cone network). Therefore, all the results valid for \Ctwo networks apply also to mixed \Cone/\Ctwo networks.

\section{Number and length of attractors}
\label{attractors}

We also performed simulations to obtain statistical properties of the
attractors of \Ctwo networks in comparison to RBNs and \Cone
networks. With the intuitive understanding developed in the previous
Section we can interpret the results and gain some new insights.

\begin{figure}
\begin{center}
\vskip 0.5cm
\includegraphics*[width=0.95\columnwidth]{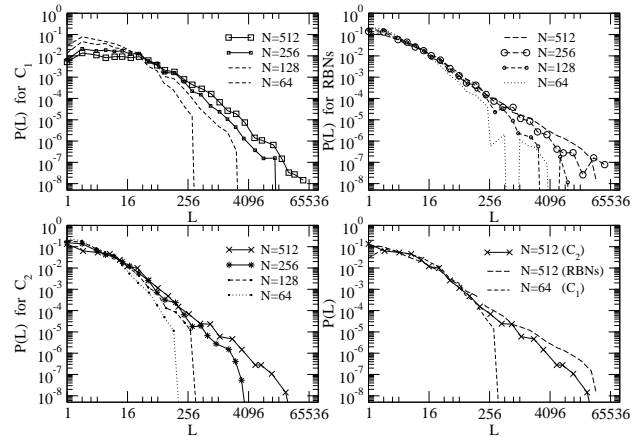}
\end{center}
\caption{Probability density distribution for the attractor lengths at different fixed network sizes $N$.
The four figures correspond respectively to \Cone networks, RBNs, canalyzing \Ctwo networks and the comparison of the three classes of networks. In the lower right figure \Cone networks with 64 nodes have on average the same number of relevant nodes as \Ctwo networks with 512 nodes. In calculating the relative frequencies of attractor lengths, the frequencies were weighted with the sizes of attractors' basins of attraction. We considered 1000 initial states for one network realization and averaged over 2000 networks. The data is binned on a logarithmic scale.}
\label{fig:alen_distr}
\end{figure}

We start with probability density for the attractor lengths. The
results are presented in Fig. \ref{fig:alen_distr}.  In order to
understand them we remind ourselves that the number of relevant nodes
$N_{rel}$ scales in \Ctwo networks and RBNs with $N^{1/3}$ , whereas
for \Cone networks it scales with $\sqrt{N}$. In all cases relevant
components of sizes less than $\sim N_{rel}$ are mainly simple
loops. The mean number of relevant loops of length $l$ is $1/l$ as
long as $l$ is sufficiently far below $N_{rel}$.  In the last graph of
Fig. \ref{fig:alen_distr} (lower right corner) the curves for $N=512$
for RBNs and \Ctwo networks agree well with the curve for \Cone
networks for $N=64$ for smaller $L$. The reason is that the number of
relevant nodes is for all three types of networks of the same size
(since $512^{1/3}=64^{1/2}$).  The difference for large $L$ is due to
the fact that in \Cone networks all relevant nodes are on simple
loops, so that all relevant components have a cycle period of the
order of their size, while the more complex components occuring in the
other network types can have much longer cycle lengths. For large $L$,
the small difference between \Ctwo networks and RBNs is due to the
presence of reversible functions in RBNs: relevant components
containing relevant nodes with two relevant inputs and a reversible
update function can have extremely long periods, which can become of
the order of the state space of the component \cite{kaufman:on}.

In our simulations, we find very long attractors also for \Ctwo
networks. This fact is fairly surprising if we remember that \Ctwo
networks were originally thought to be interesting for their short
attractors, see f.i. \cite{kauffman:metabolic}.
\begin{figure}
\begin{center}
\vskip 0.5cm
\includegraphics*[width=0.9\columnwidth]{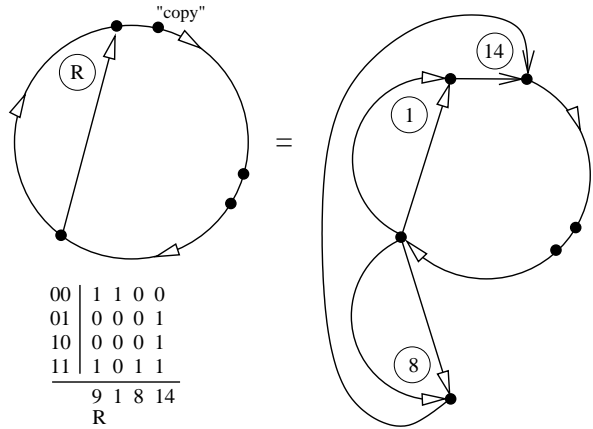}
\end{center}
\caption{An RBN relevant component and a \Ctwo relevant component,
whose attractors can be mapped pairwise onto each other.  Triangle
arrows represent a chain of nodes, and without loss of generality we
can assume that the update functions of the nodes with one input are
all a ``copy'' function.  The node marked with its ``copy'' function
is absent in the corresponding right arch of the component on the
right, and that arch is thus shorter by one node. The left arch and
the straight chain of the left component are identical to the two left
arches and the two straight chains of the right component. The table
explains the notation for the update functions, 9 being the reversible
one, which can be emulated by using the canalyzing functions 1,8, and
14. At the node with the update function $14$ the binary input combination 11
never occurs.}
\label{fig:long_attr}
\end{figure}
We can explain the appearance of very long attractors in canalyzing
networks in the following way. Let us consider a relevant component of
an RBN that is a loop with an additional chain of nodes within the loop
and a reversible update function at the node with two inputs. As shown
in \cite{kaufman:on}, the attractors of such a component can comprise
a finite proportion of the state space even for very large components. Now, it is possible to construct a
relevant component that contains three nodes with two relevant inputs, which all have a canalyzing function, and that has, for the mapping implied by Fig. \ref{fig:long_attr}, exactly the same attractor states as the relevant component of the RBN.
\comment{has exactly the
same attractor states of the nodes downstream from the (last) node with two
inputs (see Fig. \ref{fig:long_attr}). 
}
The number and lengths of
attractors is therefore identical in the two components. The
reversible function constructed from three canalyzing functions does
not, of course, appear as often in \Ctwo networks as a reversible
function occurs in the RBN. Therefore the very long
attractors appear relatively seldom in canalyzing networks.

\begin{figure}
\begin{center}
\vskip 0.5cm
\includegraphics*[width=0.95\columnwidth]{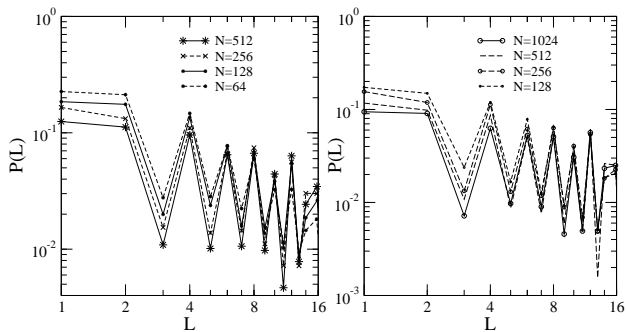}
\end{center}
\caption{Small-$L$ part of the graphs for \Ctwo networks (left) and RBNs (right) of Fig. \ref{fig:alen_distr}, without data binning, making visible even-odd oscillations.}
\label{fig:final_evenodd}
\end{figure}

In order to produce the curves in Fig. \ref{fig:alen_distr} we have
binned the data on a logarithmic scale, and we have chosen the binning
interval large enough to smoothen quite large fluctuations. Otherwise,
we could see even-odd fluctuations in the number of attractors with
neighboring lengths, see Fig. \ref{fig:final_evenodd}. There are
always more attractors with even lengths. Let us explain this
behavior. The small components of the relevant part of \Ctwo networks
are simple loops. But for simple loops even attractor lengths appear
more often, since a loop with $N_l$ nodes leads to attractors of
length $N_l$ or $2N_l$, depending on whether it is even or
odd. Therefore, an even attractor length $2N_l$ occurs in loops with
$2N_l$ or $N_l$ nodes.

The cutoff in observed attractor length is of the order $L\sim A N^2$,
with $A=0.1$ for RBN and $0.01$ for \Ctwo networks. This is a
finite size effect. Also, full ensemble averages are hard to reproduce
numerically.  For example, with a substantial probability some network
realizations appear with untypically large attractors, which lead to
an overestimation of the average attractor length for the considered
value of $N$, compare Fig. \ref{fig:avglen} below.

\begin{figure}
\begin{center}
\vskip 0.5cm
\includegraphics*[width=0.95\columnwidth]{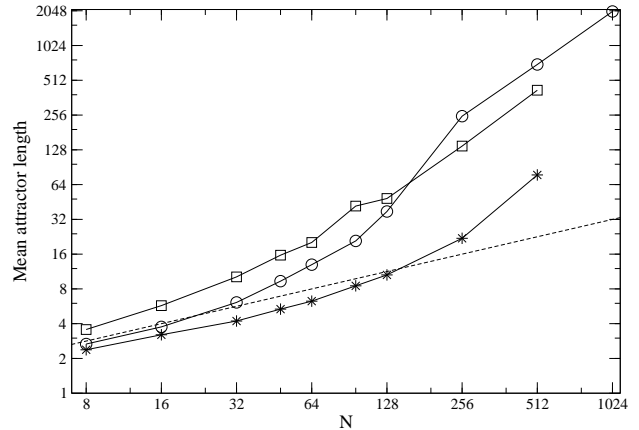}
\end{center}
\caption{Mean attractor length for canalyzing \Ctwo networks
(stars), RBNs (circles) and \Cone networks (squares) as a function of network size $N$. The dotted line is a power law $\sqrt{N}$. For the calculation of the mean attractor length, the attractor lengths of different attractors (obtained with 1000 initial states for 2000 networks) were weighted with the corresponding basins of attraction.
}
\label{fig:avglen}
\end{figure}

Fig. \ref{fig:avglen} shows our results for the mean length of
attractors.  The most important observation is that the mean attractor
length grows faster than any power law with $N$ for all the considered network
classes, including \Ctwo networks, whose attractors are not
substantially shorter than those of the other networks. Only for small
$N$ can one roughly fit the curves with the $\sqrt{N}$ law suggested a
long time ago.

All the curves are similar in shape. The reason for this is again that
the relevant part of all three network types consists mainly of simple
loops.  The mean attractor length of RBNs becomes larger than that
of the other two network classes for large $N$, due to the reversible
functions occuring in complex components. 

We now discuss simulation results for attractor numbers. In analytical calculations  \cite{samuelsson:superpolynomial,drossel:onnumber}, the average number $C_L(N_{rel})$ of attractors of length $L$ was found to scale with a power of the number of relevant nodes, 
\begin{equation}
\label{eq:C_L}
C_L(N_{rel}) \sim N_{rel}^{G_L-1}\, ,
\end{equation}
with a proportionality factor that depends on $L$. 
 $G_L$ is closely related to the number of different possible cycles of length $L$ on simple loops. One can approximately write $G_L \sim 2^L/L$, for details see \cite{drossel:onnumber}, at the end of Section $2$. 

\begin{figure}
\begin{center}
\vskip 0.5cm
\includegraphics*[width=0.9\columnwidth]{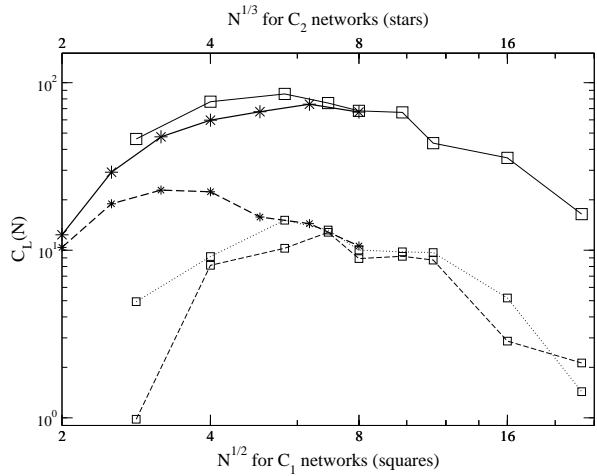}
\comment{compare samuels3.agr}
\end{center}
\caption{Absolute number of attractors of the lengths $L=7$ (dashed and dotted lines) and $L=8$ (solid lines) for canalyzing \Ctwo networks (stars) and \Cone networks (squares).
The curves for the different types of networks are
plotted in such a way that the abscissae are of the order of the number of relevant nodes for each type.
The data were extracted from those used for Fig. \ref{fig:alen_distr}.
The dotted line corresponds to the averaging over 1000 and not 2000 network realizations.
}
\label{fig:final_samuels3}
\end{figure}

Fig. \ref{fig:final_samuels3} shows our simulation results for the
number of attractors of length $L=7$ and $L=8$ for \Ctwo and \Cone
networks. The data do not contradict the theoretically predicted
scaling with $N_{rel}$ (\ref{eq:C_L}), if one realizes the limitations of computer
simulations.  We considered $1000$ initial states and could therefore
explore the state space of a network having typically of the order of
not more than 10 relevant nodes.
\comment{ In particular, attractors with a
small number of states (i.e., short attractors) cannot be found in
larger networks.}
This explains why for $N_{rel}\gtrsim 10$ the number
of found attractors decreases with $N_{rel}$ in contrast to the
analytical result. On the other hand, the analytical result is only
valid for $N_{rel} \gg L$, so that it is simply impossible to see the
predicted power law with computer simulations.  A remarkable feature
of Fig. \ref{fig:final_samuels3} is the qualitative difference between
the curves for even and odd $L$. We know from \cite{drossel:onnumber}
that $G_L$ is smaller for odd $L$ than for neighboring even $L$.
\comment{to include elsewhere:
$G_L$ shows also even-odd fluctuations. Empirical quantitative results for $G_L$
are out of reach due to large statistical errors,
compare the dotted curve in Fig. \ref{fig:final_samuels3}.
``KOMISCH'' $2^L/L$ IST GROESSER ALS 2. ``OFFENE FRAGE'' IM MOMENT\dots
}

\begin{figure}
\begin{center}
\vskip 0.5cm
\includegraphics*[width=0.95\columnwidth]{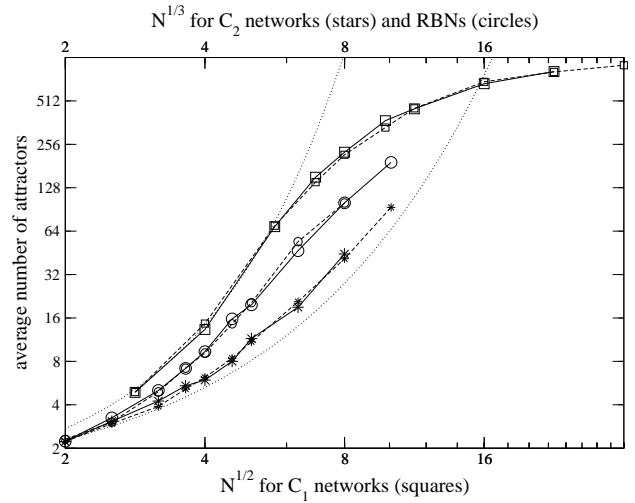}
\end{center}
\caption{Average number of attractors obtained with $1000$ initial states and $1000$ networks (dashed lines) and $2000$ networks (solid lines)
for canalyzing \Ctwo networks (stars), RBNs (circles) and \Cone networks (squares).
The curves for the different types of networks are
plotted in such a way that the abscissae are of the order of the number of relevant nodes for each type. The dotted lines correspond to the curves $2^{0.6\;\mathcal{N}_1^{0.5^{\ }}}$ and $2^{0.55\;\mathcal{N}_1^{0.7}}$, where $\mathcal{N}_1$ is the number of nodes in \Cone networks, i.e. the squared value of the abscissa here.}
\label{fig:numattr}
\end{figure}

Fig. \ref{fig:numattr} shows the total number of
attractors found starting from $1000$ initial states for each network.
The curves for the different types of networks are
plotted in such a way that the abscissae are of the order of the number of relevant nodes for each type.

Just as for the previous figure, with $\sim 2^{10}$ random initial
states the relevant nodes of a network with $N_{rel} \simeq 10$ assume a
large proportion of their possible values, and the average number of
attractors found is a good estimate of the
real ensemble average (if we average over a sufficiently large number
of network samples, compare the dashed and solid line curves in Fig. \ref{fig:numattr}).
For much larger networks, it is unlikely that we get the same
attractor twice using only $1000$ initial states. Therefore the
average number of found attractors trivially approaches $1000$,
yielding no information about the network dynamics.

We compared our results with those of \cite{socolar:scaling}, where numerical simulations were performed to obtain median number of attractors. This number is in our system less than the mean number, and the corresponding curve (not shown) lies below our data.
For \Cone networks, a lower bound for the average number of attractors is $2^{0.6\sqrt{N}}$, see \cite{drossel:number}. Our data suggest that the number of attractors can well be fitted by $2^{a N^b}$ with two constants $a$ and $b$. The lower bound as well as the fit for \Cone networks are plotted in Fig. \ref{fig:numattr} (dotted curves).

Originally based on computer simulation of small systems, Kauffman
had suggested that the mean number of attractors increases as
$\sqrt{N}$. For small $N$, our data are compatible with such a
relation.
The newer and analytical
results, see f.i. \cite{drossel:onnumber}, lead to the exponentially
large number of attractors, also in agreement with our data.

The similarity in the form of the three curves in Fig. \ref{fig:numattr} confirms
 our understanding of the dynamics in terms of the relevant nodes. In the
limit $N\to\infty$ the fraction of the relevant nodes with two inputs
goes to zero for the \Ctwo networks and RBNs and the
average number of attractors is mainly determined by of the order of $\ln N$ relevant loops. The average number of attractors grows
exponentially fast with the number of relevant nodes.

\section{Summary}

In this paper, we have shown that canalyzing random Boolean networks
have a frozen core, the size of which is comparable to that of other
random Boolean networks. It follows that the attractors of canalyzing
networks are determined by of the order of $N^{1/3}$ relevant nodes,
which are connected to of the order of $\ln N$ relevant components,
most of which are simple loops. 

We have explained how the frozen core arises starting from
self-freezing loops. Furthermore, we have investigated the numbers and
lengths of attractors. From the properties of the relevant components
it follows that their average numbers increase faster than any power
law with system size. Although attractors of canalyzing networks are
on average shorter than those of RBNs, extremely long attractors can
also arise in canalyzing networks. We have shown this by constructing
a relevant component that has the same attractors as a relevant
component of an RBN. All the results valid for \Ctwo networks
apply also to mixed \Cone/\Ctwo networks.

We have also seen that incomplete sampling leads to large fluctuations
and uncertainties in the data. Additionally, very short and very
long attractors are difficult to find. The first ones constitute an
exponentially small fraction of the state space, the others appear
exponentially seldom in a network realization.  Therefore, computer
simulations needed to be supplemented by analytical arguments. At the same
time, it is extremely difficult to verify numerically some known
analytical results.

We conclude that the original hypothesis that critical canalyzing
networks have short attractors cannot be upheld. Rather, biological
systems need to be modeled by more specific networks that are not
randomly connected.

\bibliography{ubv5}
\end{document}